\documentclass[11pt,a4paper]{article}
\usepackage[hyperref]{emnlp2018}
\usepackage{times}
\usepackage{url}
\usepackage{amsmath}
\usepackage{amssymb}
\usepackage{mathtools}
\usepackage{bm}
\usepackage{latexsym}

\usepackage{algorithm}
\usepackage[noend]{algpseudocode}
\usepackage{etoolbox}
\usepackage{float}
\usepackage{subfig}
\usepackage{multirow}
\usepackage{comment}
\usepackage{arydshln}
\usepackage{hyperref}
\aclfinalcopy 

\title{Structured Multi-Label Biomedical Text Tagging \\ via Attentive Neural Tree Decoding}

\author{Gaurav Singh \\ University College London \\ {gaurav.singh.15@ucl.ac.uk} \And
        James Thomas \\ University College London \\ {james.thomas@ucl.ac.uk } \And
        Iain J. Marshall \\ King's College London \\  {iain.marshall@kcl.ac.uk } 
        \AND
        John Shawe-Taylor \\ University College London \\  {j.shawe-taylor@ucl.ac.uk} \And
        Byron C. Wallace \\ Northeastern University \\  {b.wallace@northeastern.edu} }

\date{}

\PassOptionsToPackage{noend}{algpseudocode}
\usepackage{algpseudocode}

\errorcontextlines\maxdimen

\makeatletter

\begin{document}
\maketitle
\begin{abstract}

We propose a model for tagging unstructured texts with an arbitrary number of terms drawn from a tree-structured vocabulary (i.e., an ontology). We treat this as a special case of sequence-to-sequence learning in which the decoder begins at the root node of an ontological tree and recursively elects to expand child nodes as a function of the input text, the current node, and the latent decoder state. In our experiments the proposed method outperforms state-of-the-art approaches on the important task of automatically assigning MeSH terms to biomedical abstracts.
\end{abstract}
\section{Introduction}

We consider the task of multilabel text annotation, where labels are drawn from an ontology. We are motivated by problems in biomedical NLP \cite{zweigenbaum2007frontiers, demner2016aspiring}. Specifically, scientific abstracts in this domain are typically associated with multiple Medical Subject Heading (MeSH) terms. MeSH is a controlled, hierarchically structured vocabulary that facilitates semantic labeling of texts at varying levels of granularity. This in turn supports semantic indexing of biomedical literature, thus facilitating improved search and retrieval.\footnote{This problem also resembles tagging clinical notes with ICD codes \cite{mullenbach2018explainable}.}

At present, MeSH annotation is largely performed manually by highly skilled annotators employed by the National Library of Medicine (NLM). Automating this annotation task is thus highly desirable, and there have been considerable efforts to do so. The BIOASQ\footnote{\url{http://bioasq.org/}} challenge, in particular, concerns MeSH annotation, and competitive systems have emerged from this in past years \cite{liu2014fudan, tsoumakas2013large}; these constitute baseline approaches in the present work.

\begin{figure}
    \centering
    \includegraphics[scale=0.8]{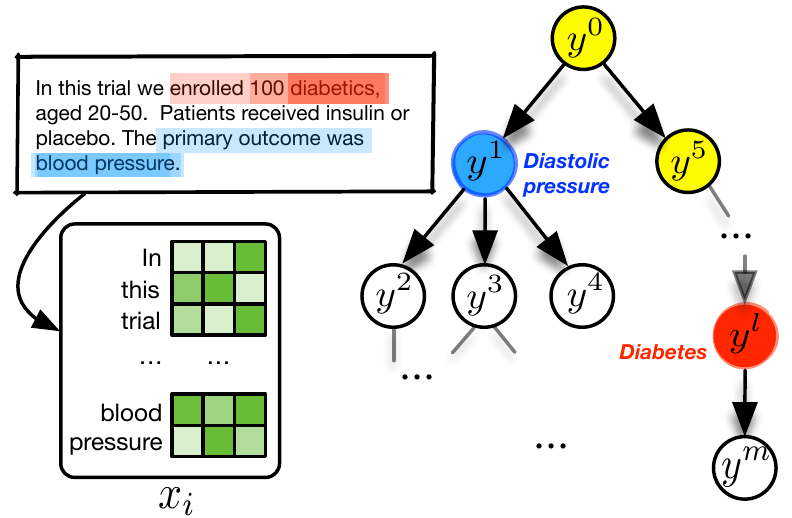}
    \caption{Illustration of the proposed Neural Tree Decoding (NTD) model. Input text is encoded, and a decoder then conditionally traverses the label tree to select all relevant nodes to apply, with node-wise attention induced over the input text.
    }
    \label{fig:our-seq2seq-model}
\end{figure}


More generally, MeSH annotation is a specific instance of multi-label classification, which has received substantial attention in general \cite{elisseeff2002kernel,furnkranz2008multilabel,read2011classifier,bhatia2015sparse,daume2017logarithmic,chen-EtAl:2017:RepL4NLP,jernite2016simultaneous}. Our work differs from these prior efforts in that MeSH tagging involves \emph{structured} multi-label classification: the label space is a tree\footnote{Technically, MeSH comprises multiple trees, but we join these by insertion of an overarching root node.} in which nodes represent nested semantic concepts, and the specificity of these increases with depth. 

Past efforts in multi-label classification have considered hierarchical and tree-based approaches for tagging \cite{jernite2016simultaneous,beygelzimer2009conditional,daume2017logarithmic}, but these have not assumed a \emph{given} structured label space; instead, these efforts have attempted to \emph{induce} trees to improve inference efficiency. By contrast, we propose to explicitly capitalize on a known output structure codified here by the target ontology from which tags are drawn. We realize this by recursively traversing the tree to make (conditional) binary tag application predictions. 


The contribution of this work is a neural sequence-to-sequence (\emph{seq2seq}) model \cite{bahdanau2014neural} for structured multi-label classification. Our approach entails encoding the input text to be tagged using an RNN, and then \emph{decoding into the ontological output space}. This involves a tree traversal beginning at the root of the tree. At each step, the decoder decides whether to `expand' children as a function of a hidden state vector, node embeddings, and induced attention weights over the input text. This approach is schematized in Figure \ref{fig:our-seq2seq-model}. Expanded nodes are added to the predicted tag set. This process is repeated recursively until either leaf nodes are reached or no children are selected for expansion. This neural tree decoding (NTD) model outperforms state-of-the-art models for MeSH tagging.

\section{Model}


{\bf Overview}. Our model is an instance of an encoder-decoder architecture. For the encoder, we adopt a standard Gated Recurrent Unit (GRU) network \cite{cho2014learning}, which yields hidden states for the tokens comprising an input document. The decoder network consumes these outputs and begins at the root of the ontological tree. It induces an attention distribution over encoder states, which is used together with the current decoder state vector to inform which (if any) of its immediate children are applicable to the input text (Figure \ref{fig:our-seq2seq-model}). This decoding process proceeds recursively for all children deemed relevant. Below we provide more in-depth technical detail regarding the constituent modules.



The encoder ($\textsc{enc}$) consumes as input a raw sequence of words, here composing an abstract. These are passed through an embedding layer, producing a sequence of word embeddings $x$ (for clarity we omit a document index here), which are then passed through a GRU \cite{cho2014properties} to obtain a sequence of hidden vectors $h = \{h_0, \cdots, h_{|x|-1}\}$, where $h_t = \text{GRU}(x_t, h_{t-1})$.


These are then passed to our neural tree decoder, which is responsible for tagging the encoded text with an arbitrary number of terms from the label tree, i.e., sequences in the structured output space. This module traverses the label space top-down, beginning at the root, thus exploiting the concept hierarchy codified by the tree structure. 

At each step in the decoding process, the decoder will be positioned at a particular node in the tree $n$. Children --- immediate descendents --- of this node are then considered for expansion in turn, based on a hidden state vector $s_n$, and a context vector $c_n$. Both of these are initialized to zero vectors and recursively updated during traversal, i.e., as nodes are selected for expansion (and hence added to the predicted tag set). More specifically, the context vector that informs the decision to expand node $v$ in the label hierarchy from its parent node $n$ is a weighted sum of the encoder hidden states $h$, where weights reflect induced attention over inputs, conditioned on $n$. That is: 



\begin{equation}
c_n = \sum_{j} \alpha_{nj} h_j
\end{equation}






\noindent where

\begin{equation}
\alpha_{nj} = \frac{\textit {exp}\{a(s_{n}, h_j)|\theta_n\}}{\sum_{l} \textit {exp}\{a(s_n, h_l)|\theta_n\}} 
\end{equation}

\noindent and $a$ is a simple multi-layer perceptron (MLP), with node-specific parameters $\theta_n$. Here both sums range over the length of the input text. 

Given $c_n$, we then estimate the probability that child label $v$ is applicable to the current input text as a function of the decoder state vector ($s_n$), the current context vector ($c_n$) and the decoder parameters. In particular, this is realized via a standard linear layer with sigmoid activations, parameterized by a weight matrix $W$ comprising independent weight vectors for each output node $v$. Thus the score for a particular output node $v$ is $\sigma(W_v \cdot \bm{[}s_n, c_n\bm{]})$, where $W_v$ denotes the weight vector for output node $v$.

\algrenewcommand{\algorithmicrequire}{\textbf{Input:}}
\algrenewcommand{\algorithmicensure}{\textbf{Output:}}
\algrenewcommand{\algorithmicforall}{\textbf{for each}}
\newcommand{\To}{\textbf{to}\xspace}
\newcommand{\Dosth}{\State \textbf{Do Something}\xspace}
\newcommand{\Please}[1]{\State \textbf{#1}}
\newcommand{\ForEach}{\ForAll}

\begin{algorithm}
\small
    \label{algo} 
    \caption{\textsc{RecursiveTreeDecoding} }
    \begin{algorithmic}[1]
        \Function{NodeLoss}{$n$, $h$, $s$, $y$}
        \State $l_n$ $\gets$ 0
        \State $c_n$, $s_n$ $\gets$ \textsc{dec}($h$, $n$, $s$)
        
         \ForEach{child $v \in$ children($n$)}
          \State $\hat{y}_v$ $\gets$ $\sigma(W_v \cdot \bm{[}s_n, c_n\bm{]}$)
          \State $p_v \gets$ $\propto$ depth in tree
          \State $B_v \sim$  Ber($p_v$)
          \If {$B_v$}
            \State $l_n$ $\gets l_n + \mathcal{L}(\hat{y}_v, y)$
          \EndIf
          
          \If{$\hat{y}_v > \tau$} 
             \State $l_n$ $\gets$ $l_n$ + \textsc{NodeLoss}($v$, $h$, $s_n$, $y$)
          \EndIf
         \EndFor
        \Return $l_n$
        \EndFunction
        \vspace{0.5em}
        \Function{Train}{$x$, $y$, $\alpha$, epochs}
        \State $\theta \gets $ \textsc{init}$(\theta)$
        \State $e \gets 0$
        \While{$e <$ epochs}
            \ForAll{instance $x_i \in x$}
                \State $h_i \gets $ \textsc{enc}($x_i$)
                \State $s_0 \gets {\mathbf 0}$
                \State $l_i \gets$ \textsc{NodeLoss}(\textsc{root}, $h_i$, $s_0$, $y_i$)
                \State $\Delta \theta \gets$ \textsc{backprop}($l_i$) 
                \State $\theta \gets \theta + \alpha \Delta \theta$
            \EndFor
            \State $e \gets e + 1$
        \EndWhile
        \Return $\theta$
        \EndFunction
    \end{algorithmic}
    \label{alg:train}
\end{algorithm}

Pseudocode for the training and decoding procedures are presented in Algorithm \ref{alg:train}. In the $\textsc{NodeLoss}$ function, $n$ denotes a particular node. The set of hidden vectors induced by the encoder (corresponding to the inputs) are denoted by $h$, $s$ is the hidden state of the decoder, and $y$ is the reference label (this encodes a path in the output tree). We assume the decoder, $\textsc{Dec}$, consumes input representations, a node index and a hidden state and yields a context vector for $n$, $c_n$ and an updated state vector $s_n$; in our case the latter is implemented via a GRU. The advantage of using an RNN during decoding is that this allows the exploitation of learned, distributed hidden representations of partial tree paths, which inform node-wise attention and subsequent predictions.

Incurring loss for all nodes along the path specified by $y$ would place a disproportionate amount of emphasis on correctly applying terms that are `higher' in the ontology, as loss will be propagated for the initial predictions concerning the application of these and then also, due to recursive application, for all of their children (and so on). Thus we only incur (and hence backpropagate) loss for a node $v$ stochastically, according to a Bernoulli distribution $B$ with parameter $p_v$. We set $p_v$ to be proportional to the depth of node $v$ in the tree such that we are likely to incur larger loss for deeper (rarely occurring) nodes. We operationalize this as: $p_v = {\textit{min}}(1, 0.5 + \frac{m}{f_v})$, where $m$ is the count corresponding to the least frequently observed node in the training corpus and $f_v$ is the count for node $v$. In Section \ref{section:results} we demonstrate the benefit of this approach.

At train time we use \emph{teacher forcing} \cite{williams1989learning} during decoding. That is, we revert the model back to the correct (training) tree subsequence when it goes off-course, and continue decoding from there. We have elided this detail from the pseudocode for clarity.

\section{Experimental setup}

Below we describe experimental details concerning our implementation, datasets and baselines. Code and data to reproduce our results is available at \url{https://github.com/gauravsc/NTD}. 

\subsection{Implementation Details}
 We limited the vocabulary to the $50,000$ most frequent words. Word embeddings were initialized to pre-trained vectors induced via word2vec, trained over a large set of abstracts indexed on PubMed.\footnote{A repository of biomedical literature.} Ontology node embeddings were pre-trained using DeepWalk \cite{perozzi2014deepwalk}, fit over PubMed. 

\subsection{Dataset}
Our dataset comprises abstracts of articles describing randomized controlled trials (RCTs) from PubMed along with their MeSH terms. The MeSH annotations were manually applied by professionals at the National Library of Medicine (NLM). The label space underlying MeSH terms is codified by a publicly available ontology.\footnote{\small \url{https://meshb-prev.nlm.nih.gov/treeView}} 

We split this dataset into disjoint sets for training/development and final evaluation (Table \ref{tab:dataset}). We further separated the former into train, validation and development test subsets, to refine our approach. For our final evaluation we used a heldout set of 10,000 abstracts that were not seen in any way during model development and/or hyperparameter tuning. We performed extensive hyperparameter tuning for the baseline models to ensure fair comparison; details regarding this tuning are provided in the Appendix.

\subsection{Baselines}
We compare our proposed approach to three baselines, including two prior winners of the annual BioASQ challenge, which includes an automated MeSH annotation task. However, it is important to note that we used a different (and considerably smaller) dataset in the current work, as compared to the corpus used in the BioASQ challenge. 

\vspace{.25em}

\noindent \textbf{LSSI} \cite{tsoumakas2013large} use an approach that involves predicting both the number of terms and which to apply to a given abstract. They use linear models for both tasks, which operate over TF-IDF representations of abstracts. Specifically, they train a regressor to predict $k$, the number of MeSH terms to be applied to an abstract. Simultaneously, a binary linear SVM is trained independently for each MeSH term appearing in the train set. At test time, these SVMs provide scores for each term and the top $\hat{k}$ terms are applied, where $\hat{k}$ is the estimate from the aforementioned regressor. 

\vspace{.25em}

\noindent \textbf{UIUC} \cite{liu2014fudan} uses a learning-to-rank model to identify the top MeSH terms for an abstract from a candidate set of terms, which is obtained from the nearest neighbours of the abstract. Additionally, one SVM classifier is trained for each of the MeSH terms (similar to the above approach), and scores for each are used to obtain additional terms to be added to the candidate set. In the end, a threshold (tuned on the validation set) is used to select the final set of terms to be assigned.

\vspace{.25em}

\noindent Finally, we consider a deep multilabel classification model \textbf{DML} \cite{rios2015convolutional} that takes as input unstructured abstracts and activates the output nodes corresponding to the relevant MeSH terms. In brief, embedded tokens are fed through a CNN to induce a vector representation, which is then passed on to the dense output layer. Finally, this is passed through a sigmoid activation function. Note that this model exploits the same pre-trained word embeddings as our model does.

\begin{table}
    \centering
    \begin{tabular}{l l}
         Train &  20000 \\ 
         Validation & 4000\\
         Dev test & 18884\\ 
         Test (held-out) & 10000 \\
         \hline
         Mean MeSH terms per article & 15.33\\ 
         Total unique MeSH terms & 27892 \\
         Unique MeSH terms in dataset & 3781\\ 
    \end{tabular}
    \caption{Dataset statistics.}
    \label{tab:dataset}
\end{table}

\subsection{Evaluation metrics}
\label{sec:eval}
We first evaluate model performance via output node-wise precision, recall and F1 measure. However, these metrics are overly strict in the sense that a model will be penalized equally for all mistakes, regardless of whether they are nearby or far from the target in the label tree. This is problematic because whether to apply a specific MeSH term or its immediate parent may be somewhat subjective in practice. To quantify this, and to explore the extent to which explicitly decoding into the target label space yields improved predictions, we also consider a measure that we refer to as \emph{semantic distance} (SD):


\begin{equation}
\text{SD} = \frac{1}{|\mathcal{Y}|}\sum_{u\in \mathcal{Y}} \min_{v \in \mathcal{\hat{Y}}}{\textit{dist}}(u, v)
\label{equation:sp}
\end{equation}
    
\noindent where $\mathcal{Y}$ and $\mathcal{\hat{Y}}$ are the sets of target and predicted terms respectively, and ${\textit{dist}}$ is a function that returns the shortest distance between two nodes in the label ontology tree. The idea is that this penalizes less for `near misses'. Thus if a model fails to apply a particular tag $t$, but does apply one near to $t$ in the label tree, then it is penalized less.\footnote{This metric is equivalent to the sum of two metrics ("divergent path to gold standard" and "divergent path to prediction") defined in \cite{perotte2013diagnosis}.} We hypothesize that our model will improve results markedly with respect to this metric, given our exploitation of the tree structure. 

As in the case of recall, SD can be `gamed': one can achieve a perfect score by predicting that all nodes apply to a given abstract. Thus this is only meaningful alongside complementary metrics like F1. 
\section{Results}
\label{section:results}

Results on the test set (which was completely held out during development) are reported in Table \ref{tab:res3}. The proposed Neural Tree Decoding model with stochastic backpropagation (NTD-s) bests the most competitive baseline (LSSI) in F1 score by over 2 points. 

To explore the effect of backpropagating loss from nodes in proportion to their depth in the ontology, we also include results for a deterministic variant that does not do this, NTD-d. This version does not perform as well, demonstrating the utility of the proposed training approach.

The metrics reported thus far do not account of the structure in the output space. We thus additionally report results with respect to the the semantic distance (SD) metric (Eq. \ref{equation:sp}). We observe a marked performance increase of $\sim$21$\%$ over the best performing baseline. This is intuitive given that we are explicitly decoding into the label tree structure, and demonstrates the ability of our model to learn the ontological structure, thereby predicting semantically appropriate terms.



\begin{table}
    \centering
    \begin{tabular}{c | c c c  c }
    \hline
         Method & Precision & Recall & F1 & SD \\ \hline
         LSSI & 0.326 & 0.293 & 0.309  & 1.518 \\ 
         UIUC & 0.236 & {\bf 0.388} & 0.291 & 1.433 \\ 
         DML & 0.378  & 0.223  & 0.275 &  1.516\\ \hline
         NTD-d & {\bf 0.434} &  0.235 & 0.299 & 1.209 \\
         NTD-s &  0.425 & 0.265 & {\bf 0.327} & \textbf{1.130}\\
    \end{tabular}
    \caption{Results on the held-out test dataset. SD refers to \emph{semantic distance}, defined in Eq. \ref{equation:sp}.}
    \label{tab:res3}
\end{table}

\section{Conclusions, Discussion \& Limitations}

We developed a neural attentive sequence tree decoding model for structured multilabel classification where labels are drawn from a known ontology. The proposed method can decode an input text into a tree of labels, effectively using the structure in the output space. We demonstrated that this model outperformed SOTA approaches for the important task of tagging biomedical abstracts with Medical Subject Heading (MeSH) terms on a modestly sized training corpus. Code and data to reproduce these results are available at \url{https://github.com/gauravsc/NTD}.

One limitation of our model is that it is comparatively slow, due to having to traverse the tree structure during decoding. Prediction speed may not be a major issue in practice, as articles on PubMed could be batch tagged nightly as they arrive. However, slow decoding also means lengthy training (see Appendix, section A.2 for details). For this reason we have here used a modest training set of $\sim$20k abstracts, which is smaller than corpora used in prior work on this task. Given the relative expressiveness of our model, we expect it to benefit substantially from additional training data, moreso than the simpler baseline architectures. But at present this is only a conjecture.

In future work we thus hope to apply this model to larger datasets, and to address the efficiency issue. Concerning the latter, sibling subtrees may be traversed in parallel, conditioned on the hidden state of their parent. Another promising direction would be to move to convolutional encoder and decoder architectures, designing the latter in a way similarly capitalizes on the label space tree structure.



\section{Acknowledgements}
JT and GS acknowledge support from Cochrane via the Transform project. BCW was supported by the National Library of Medicine (NLM) of the National Institutes of Health (NIH), grant R01LM012086. IJM acknowledges support from the MRC (UK), through its grant MR/N015185/1.
\bibliography{naaclhlt2018}
\bibliographystyle{acl_natbib_nourl}
\appendix
\section{Appendix}
\subsection{Parameter Tuning}
We performed extensive parameter tuning for the baselines. For LSSI, we tuned over vocabulary sizes of 50K, 100K and 150K. We also tuned over the regularization parameter used for linear-SVR and linear-SVM on a validation set. The values were tuned over a range of [0.1, 1.0, 10.0, 100.0]. For UIUC, we tuned over the vocabulary sizes of 50K, 100K and 150K. We also tuned the $k$NN classifier for $k$ $\in$ [5, 10, 15, 30]. We also tuned the regularizer of the linearSVM in the range [0.1, 1.0, 10.0, 100.0]. Afterwards, we tuned the threshold of the classifier for ten equidistant values in the range $(0, 1)$. For DML, we performed nested validation  after each epoch to save the best performing model parameters. We also tuned the threshold for classification over ten equidistant values in the interval $(0,1)$.  
\subsection{Training Details}
The training times for iterating over 1K samples can range in 30-40 minutes on Tesla K60 GPU, although, it can be faster with more advanced GPUs. We can converge with a learn rate of 0.01 in 30-40 epochs. We can converge over a set of 20K documents in $\approx$ 3-4 days.

\end{document}